\documentclass[aps,prb,showpacs,floatfix,twocolumn,byrevtex,superscriptaddress]{revtex4-1}
\usepackage[english]{babel}
\usepackage{amsmath}
\usepackage{graphicx}
\usepackage{float}
\usepackage{bm}
\usepackage{multirow}
\usepackage{dcolumn}
\usepackage{hyperref}
\newcommand{\icm}{\ensuremath{~\textrm{cm}^{-1}}}% % cm-1
\newcommand{\BKFA}{Ba$_{0.6}$K$_{0.4}$Fe$_{2}$As$_{2}$}

\newcommand{\BFA}{BaFe$_{2}$As$_{2}$}
\newcommand{\BFAP}{BaFe$_{2}$(As$_{0.85}$P$_{0.15}$)$_{2}$}

\newcommand{\BFCA}{Ba(Fe$_{0.92}$Co$_{0.08}$)$_{2}$As$_{2}$}

\begin{document}

%%%%%%%%%%%%%%%%%%%%%%%%%%%%%%%%%%%%%%%%%%%%%%%%%%%%
%
%   Title, authors and affiliations
%

\title{\boldmath Anomalous phonon red shift in K-doped BaFe$_{2}$As$_{2}$ Iron-pnictides \unboldmath}
\author{B. Xu}
\affiliation{LPEM, ESPCI-ParisTech, PSL Research University, 10 rue Vauquelin, F-75231 Paris Cedex 5, France}
\affiliation{Beijing National Laboratory for Condensed Matter Physics, Institute of Physics, Chinese Academy of Sciences, P.O. Box 603, Beijing 100190, China}
\author{Y. M. Dai}
\email[]{ymdai@lanl.gov}
\affiliation{Condensed Matter Physics and Materials Science Department, Brookhaven National Laboratory, Upton, New York 11973, USA}
\author{B. Shen}
\author{H. Xiao}
\affiliation{Beijing National Laboratory for Condensed Matter Physics, Institute of Physics, Chinese Academy of Sciences, P.O. Box 603, Beijing 100190, China}
\author{Z. R. Ye}
\affiliation{State Key Laboratory of Surface Physics, Department of Physics, and Advanced Materials Laboratory, Fudan University, Shanghai 200433, China}
\author{A. Forget}
\author{D. Colson}
\affiliation{IRAMIS, SPEC, CEA, 91191 Gif sur Yvette, France}
\author{D. L. Feng}
\affiliation{State Key Laboratory of Surface Physics, Department of Physics, and Advanced Materials Laboratory, Fudan University, Shanghai 200433, China}
\author{H. H. Wen}
\affiliation{National Laboratory of Solid State Microstructures and Department of Physics, Nanjing University, Nanjing 210093, China}
\author{C. C. Homes}
\affiliation{Condensed Matter Physics and Materials Science Department, Brookhaven National Laboratory, Upton, New York 11973, USA}
\author{X. G. Qiu}
\affiliation{Beijing National Laboratory for Condensed Matter Physics, Institute of Physics, Chinese Academy of Sciences, P.O. Box 603, Beijing 100190, China}
\author{R. P. S. M. Lobo}
\email[]{lobo@espci.fr}
\affiliation{LPEM, ESPCI-ParisTech, PSL Research University, 10 rue Vauquelin, F-75231 Paris Cedex 5, France}
\affiliation{CNRS, UMR 8213, F-75005 Paris, France}
\affiliation{Sorbonne Universit\'es, UPMC Paris 6, F-75005 Paris, France}

\date{\today}

%%%%%%%%%%%%%%%%%%%%%%%%%%%%%%%%%%%%
%
% Abstract
%

\begin{abstract}
The effect of K-, Co- and P-doping on the lattice dynamics in the BaFe$_2$As$_2$ system is studied by infrared spectroscopy. We focus on the phonon at $\sim$ 253~cm$^{-1}$, the highest energy in-plane infrared-active Fe-As mode in BaFe$_2$As$_2$. Our studies show that the Co- and P-doping lead to a blue shift of this phonon in frequency, which can be simply interpreted by the change of lattice parameters induced by doping. In sharp contrast, an unusual red shift of the same mode was observed in the K-doped compound, at odds with the above explanation. This anomalous behavior in K-doped BaFe$_2$As$_2$ is more likely associated with the coupling between lattice vibrations and other channels, such as charge or spin. This coupling scenario is also supported by the asymmetric line shape and intensity growth of the phonon in the K-doped compound.
\end{abstract}

%%%%%%%%%%%%%%%%%%%%%%%%%%%%%%%%%%%%%%%%
%
%   PACS numbers
%

%  74.25.Gz	Optical properties
%  78.30.-j Infrared and Raman spectra
%  74.25.Kc	Phonons
%  72.80.-r	Conductivity of specific materials


\maketitle

%%%%%%%%%%%%%%%%%%%%%%%%%%%%%%%%%%%%%%%%%%%%%%%%%%%%%%%%%%%%%%%%%%%%%%%%%%%%%%%
%
% Introduction
%

In spite of extensive studies on high-$T_{c}$ superconductivity in iron pnictides since its discovery,\cite{Kamihara2008} the question of what plays an important role in the paring mechanism of this class of superconductors remains enigmatic. Theoretical calculations have demonstrated that, unlike the traditional BCS superconductors, an electron-phonon interaction is not sufficient to account for such a high $T_c$ in iron pnictides.\cite{Boeri2008} Therefore, the $s_{\pm}$ paring with a sign reversal in the gap function, that is mediated by spin fluctuations, was proposed by Mazin \emph{et al.}.\cite{Mazin2008} However, the above scenario seems far from being thoroughly established. The $s_{\pm}$ paring state is expected to be sensitive to impurity scattering.\cite{Onari2009} Indeed, on one hand in- and out-of-plane dopings affect $T_c$ and the unpaired quasiparticle density in the \BFA\ family,\cite{Dai2013a} but on the other hand, superconductivity is robust against impurities in 1111 materials.\cite{Lee2010,Tarantini2010} Furthermore, a large iron isotope effect has been reported in SmFeAsO$_{1-x}$Fe$_{x}$ and Ba$_{1-x}$K$_{x}$Fe$_{2}$As$_{2}$,\cite{Liu2009} indicating that electron-phonon coupling plays some role in the paring mechanism. By taking these facts into account, Kontani \emph{et al.} proposed that electron-phonon coupling arising from the Fe-ion oscillation can induce orbital fluctuations, mediating the $s_{++}$ paring without sign reversal.\cite{Kontani2010} A recent quantitative convergent beam electron diffraction study by Ma \emph{et al.}\cite{Ma2014} has revealed strong coupling between Fe orbital fluctuations and anion-dipole polarizations in Ba(Fe$_{1-x}$Co$_{x}$)$_{2}$As$_{2}$. They suggest that a full understanding of the paring mechanism in iron-pnictides can only be reached by considering the charge, spin, orbital, lattice and anion polarization all together in a consistent theory.

Infrared spectroscopy is a standard tool to investigate lattice dynamics, providing information on the coupling between lattice vibrations and electrons or spins.\cite{Choi2003,Padilla2005,Kuzmenko2009,Xu2014} Although optical investigations into the lattice dynamics of iron pnictides have been conducted by many groups,\cite{Akrap2009,Schafgans2011,Charnukha2013,Xu2014} a comparison study of different iron pnictides, in particular different substitution types in the same family, has never been performed.

We fill this gap by comparing the behavior of the $\sim$253~\icm\ in-plane infrared-active Fe-As mode in BaFe$_{2}$As$_{2}$ (parent compound of the Ba122 family) and three different doping types (K, Co and P doping). We observe a blue shift of this mode in the Co- and P-doped BaFe$_{2}$As$_{2}$, as well as an anomalous red shift of the same mode in the K-doped compound. The latter can not be explained by the change of lattice parameters induced by doping. A close inspection of the phonon line shape and intensity leads us to the conclusion that the coupling between lattice vibrations and other channels is stronger in K-doped BaFe$_{2}$As$_{2}$ and thus responsible for the anomalous red shift of the phonon.

%%%%%%%%%%%%%%%%%%%%%%%%%%%%%%%%%%%%%%%%%%%%%%%%%%%%%%%%%%%%%%%%%%%%%%%%%%%%%%%
%
% Experiments
%

High quality single crystals of \BFA\ (BFA, $T_{N}$ $\simeq$ 138 K), \BKFA\ (K40, $T_{c}$ $\simeq$ 39 K), and \BFCA\ (Co08, $T_{c}$ $\simeq$ 23 K) were grown with a self-flux method,\cite{Shen2011,Rullier-Albenque2009} while the \BFAP\ (P15, $T_{N}$ $\simeq$ 90 K) single crystals were grown without flux.\cite{Ye2012} The \emph{ab}-plane reflectivity $R(\omega)$ was measured at a near-normal angle of incidence on Bruker IFS113v and IFS66v spectrometers. An \emph{in situ} gold overfilling technique\cite{Homes1993} was used to obtain the absolute reflectivity of the samples. Data from 30 to $15\,000\icm$ were collected at different temperatures on freshly cleaved surfaces for each sample, and then we extended the reflectivity to $40\,000\icm$ at room temperature with an AvaSpec-2048 $\times$ 14 optical fiber spectrometer.

%%%%%%%%%%%%%%%%%%%%%%%%%%%%%%%%%%%%%%%%%%%%%%%%%%%%%%%%%%%%%%%%%%%%%%%%%%%%%%%
%%
%% Data analysis
%%

%%%%%%%%%%%%%%%%%%%%%%%%%%%%%%%%%%%%%%%%%%
%%
%%  Reflectivity
%%

%%%%%%%%%%%%%%%%%%%%%%%%%%%%%%%%%
%Figure 1
\begin{figure}[tb]
\includegraphics[width=0.9\columnwidth]{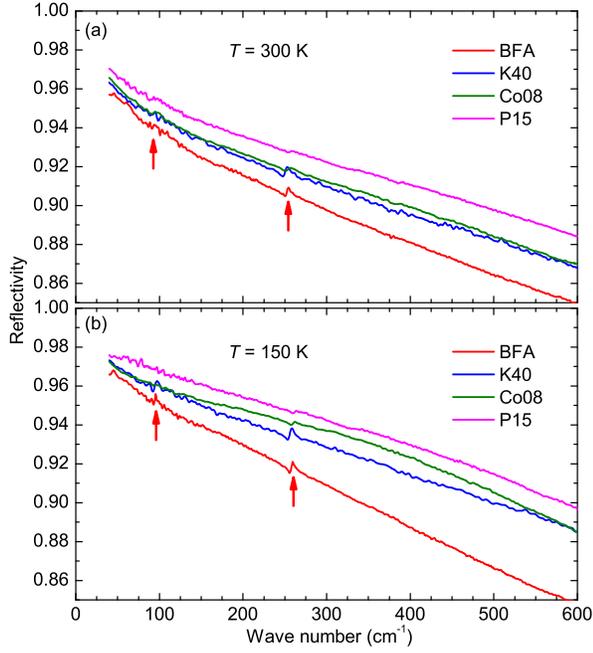}
\caption{ (color online) The reflectivity in the far infrared region for BFA, K40, Co08 and P15 at 300~K (a) and 150~K (b). The red arrows indicate the $ab$-plane infrared-active Ba mode at 94\icm\ and Fe-As mode at 253~\icm\ in BFA.}
\label{Fig1}
\end{figure}
Figure~\ref{Fig1}(a) shows the far-infrared reflectivity at room temperature (300~K) for all 4 compounds. Metallic behavior can be realized in all these materials by their relatively high reflectivity that approaches unity at zero frequency. In addition, two sharp features (indicated by the arrows), representing the symmetry-allowed $ab$ plane infrared-active $E_u$ modes, were observed at $\simeq$ 94~\icm\ and 253~\icm\ in BFA, consistent with previous works.\cite{Akrap2009,Schafgans2011} The 253~\icm\ mode is also clearly observed on the room temperature $R(\omega)$ in the doped BFA compounds. Figure~\ref{Fig1}(b) displays $R(\omega)$ for the same materials at 150~K, where similar features are revealed.

Here we would like to point out that BFA exhibits structural and magnetic phase transitions at 140~K\cite{Rotter2008a} with a consequent renormalization of the infrared phonon spectra,\cite{Akrap2009,Schafgans2011,Charnukha2013} whereas such transitions are either suppressed or absent in the doped compounds, meaning that below 140~K, these compounds are in different phases. In order to avoid effects related to these phase transitions, the temperature window in our study is constrained between 150~K and 300~K.

%%%%%%%%%%%%%%%%%%%%%%%%%%%%%%%%%%%%%%%%%%%%%%%%
%%
%%  KK analysis and Optical conductivity
%%

In the following, we concentrate on the 253~\icm\ mode, which involves the displacements of Fe and As atoms.\cite{Akrap2009,Schafgans2011} To investigate the doping effect on this mode in a straightforward way, we calculated the optical conductivity via Kramers-Kronig analysis of the reflectivity. At low frequency, we employed a Hagen-Rubens ($R = 1 - A\sqrt{\omega}$) extrapolation. Above $40\,000\icm$ (the highest measured frequency), we utilized a constant reflectivity up to 12.5~eV, followed by a free-electron ($\omega^{-4}$) response.

%%%%%%%%%%%%%%%%%%%%%%%%%%%%%%%%%
%Figure 2
\begin{figure}[tb]
\includegraphics[width=0.9\columnwidth]{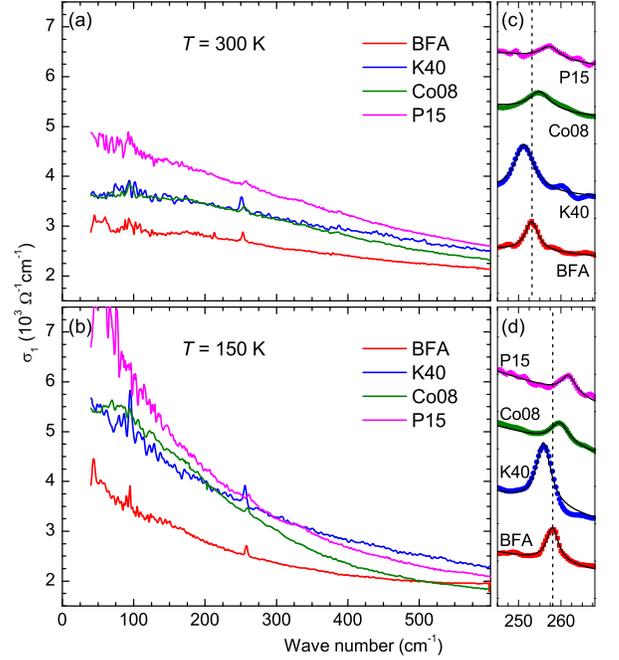}
\caption{ (color online) The low frequency optical conductivity $\sigma_{1}(\omega)$ for BFA, K40, Co08 and P15 at 300~K (a) and 150~K (b). Panel (c) and (d) show the enlarged view of Panel (a) and (b), respectively, focusing on the phonon at 253~\icm (with offset). The thin black lines through the data are the Lorentzian fitting results. The vertical dashed lines indicate the phonon frequency for BFA.}
\label{Fig2}
\end{figure}
Figure~\ref{Fig2}(a) and Fig.~\ref{Fig2}(b) show the real part of the optical conductivity $\sigma_1(\omega)$ in the far-infrared region for all compounds at 300~K and 150~K, respectively. The low-frequency $\sigma_{1}(\omega)$ exhibits a prominent Drude-like metallic behavior for all materials at both 300~K and 150~K, consistent with the reflectivity analysis.

The optical conductivity can be conveniently parameterized by a Drude-Lorentz model:
\begin{equation}
\label{DrudeLorentz}
\sigma_{1}(\omega)=\frac{2\pi}{Z_{0}}\biggl[\sum_{k}\frac{\Omega^{2}_{p,j}}{\omega^{2}\tau_j + \frac{1}{\tau_j}} + \sum_{k}\frac{\gamma_{k}\omega^{2}\Omega^{2}_{k}}{(\omega^{2}_{0,k} - \omega^{2})^{2} + \gamma^{2}_{k}\omega^{2}}\biggr],
\end{equation}
where $Z_{0}$ is the vacuum impedance. The first term describes a sum of free-carrier Drude responses, each characterized by a plasma frequency $\Omega_{p,j}$ and a scattering rate $1/\tau_j$. The second term is a sum of Lorentz oscillators, each having a resonance frequency $\omega_{0,k}$, a line width $\gamma_k$ and an oscillator strength $\Omega_k$. The optical response of FeSCs can be modeled reasonably well by the superposition of two Drude components and a series of Lorentz terms over a wide frequency range, which has been discussed in detail in previous works.\cite{Wu2010,Tu2010,Nakajima2010,Dai2013b}

In addition to the gross features, the 253~\icm\ mode manifests itself as a sharp peak in the optical conductivity. Figure~\ref{Fig2}(c) highlights the region around the 253~\icm\ phonon for all compounds. The dashed line denotes the phonon peak position of BFA. It can be immediately noticed that the phonon in the K-doped Ba122 compound shifts to lower frequency (red shift or softening), while P- and Co-doping lead to a shift of this mode to higher frequency (blue shift or hardening). Exactly the same behavior is observed at 150~K, as shown in Fig.~\ref{Fig2}(d).

In order to quantitatively analyze the behavior of the phonon upon doping, we fit it to a Lorentz oscillator with a linear background in a narrow frequency range centered at the phonon resonance frequency for all the materials at all measured temperatures. The fitting results at 300~K and 150~K are shown as thin solid lines through the corresponding data in Fig.~\ref{Fig2}(c) and Fig.~\ref{Fig2}(d), respectively. The fitting parameters are summarized in Table~\ref{Table1}.

%%%%%%%%%%%%%%%%%%%%%%%%%%%%%%%%%
%Figure 3
\begin{figure}[tb]
\includegraphics[width=0.9\columnwidth]{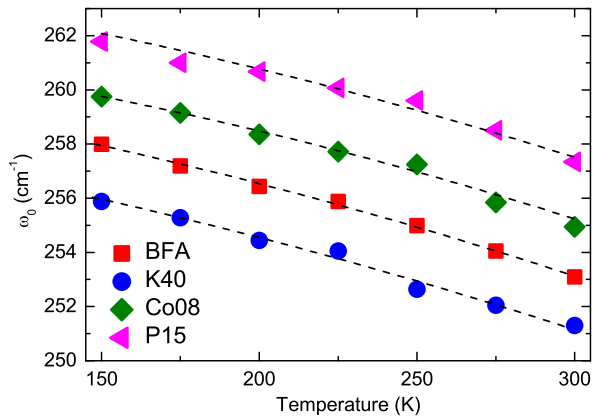}
\caption{ (color online) The resonance frequency $\omega_{0}$ of the Fe-As mode as a function of temperature for all the compounds. The black dashed lines represent a quadratic temperature dependence.}
\label{Fig3}
\end{figure}
Figure~\ref{Fig3} shows the phonon resonance frequency $\omega_{0}$ determined from the fit for all the compounds at 7 measured temperatures between 150~K and 300~K. For each material, $\omega_{0}$ increases upon cooling, following a quadratic $T$ dependence (indicated by the black dashed lines), which is expected in the absence of structural or magnetic transitions. The $T$ dependence of this phonon is in good agreement with previous works.\cite{Akrap2009} For all the measured temperatures, with respect to BFA (red solid squares), $\omega_{0}$ is smaller in K-doped compounds (blue solid circles), but larger in the Co- (green solid diamonds) and P-doped (pink solid triangles) materials.

%%%%%%%%%%%%%%%%%%%%%%%%%%%%%%%%%%%%%%%%%%%%%%%%%%%%%%%%%%%
%%
%% Discussion
%%

\begin{figure}[tb]
\includegraphics[width=0.9\columnwidth]{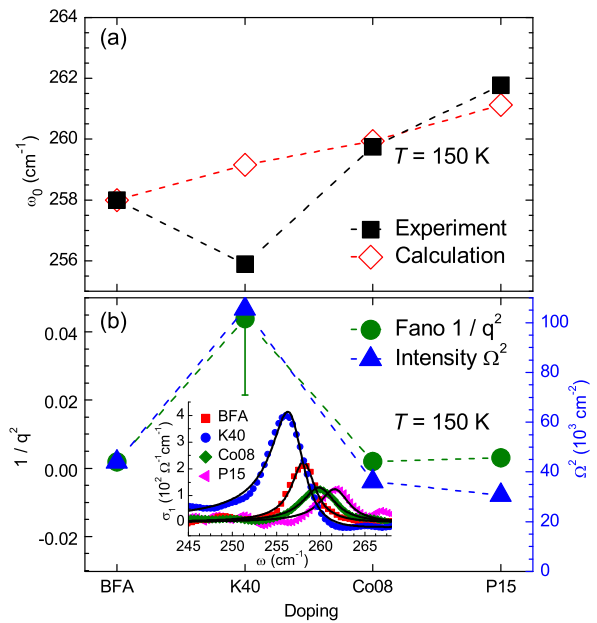}
\caption{ (color online) (a) The black solid squares denote the resonance frequency $\omega_{0}$ of the Fe-As mode at 150~K for BFA, K40, Co08, and P15. The red open diamonds are the calculated frequency by considering the Fe-As bond length change only for each compound. (b) The green solid circles and blue solid triangles portray the Fano parameter $1/q^2$ and phonon intensity $\Omega^{2}$ determined from the Fano fit at 150~K, respectively. Inset: the phonon line shape without the electronic background for all the compounds. The solid lines through the data are the Fano fitting results.}
\label{Fig4}
\end{figure}

We now trace the origin of the phonon frequency shift induced by different doping types. Doping usually has three main effects on materials: (i) changing the carrier concentration by adding electrons (electron doping) or holes (hole doping); (ii) introducing disorder; (iii) application of chemical pressure that can change the lattice parameters. Increasing the carrier concentration will increase the electronic background, potentially leading to a growth in the screening effect. However, the screening effect only affects the intensity of the phonon, but has no influence on the frequency.\cite{Homes2000} In addition, the screening effect on the ~253~\icm\ mode in the Ba122 system has been demonstrated to be very weak or absent.\cite{Akrap2009} Therefore, we can safely rule out this possibility. Disorder in the Fe-As layers may reduce the intensity of this mode, but virtually has no effect on the phonon frequency either. Finally, the change of lattice parameters due to chemical pressure may play an important role in shifting the phonon frequency. Generally, the frequency shift of a phonon is closely related to the change of the bond length $l$ of the associated atoms. The phonon shifts toward higher frequency if $l$ shrinks. The expected frequency shift can be estimated via a simple formula $\omega_{0}/\omega_{0}^{\prime}$ = $(l^{\prime}/l)^{3/2}$.\cite{Hadjiev2008} By comparison with BFA, the Fe-As bond length $l$ is 0.3\% shorter in K40,\cite{Rotter2008a,Rotter2008c,Rotter2008b} 0.5\% shorter in Co08\cite{Drotziger2010} and 0.8\% shorter in P15.\cite{Rotter2010} Based on these parameters, the doping-induced frequency shift of the 253~\icm\ mode in BFA can be easily estimated in all three compounds. The open diamonds in Fig.~\ref{Fig4}(a) denote the calculated phonon frequency by considering the Fe-As bond length for each material, which we compare to the values determined from the experiment (solid squares). Note that the measured phonon frequency agrees very well with the calculation in both Co08 and P15, suggesting that the phonon frequency shift is dominated by the Fe-As bond length. In sharp contrast, a striking difference between the calculated phonon frequency and the experimental result is found in K40. K doping induces a contraction of the Fe-As bond and, hence, one expects a blue shift of the corresponding phonon. However, a prominent red shift of the phonon frequency is observed from the experiment, which can not be explained by the change of the Fe-As bond length alone.

%%%%%%%%%%%%%%%%%%%%%%%%%%%%%%%%%
%Table 1
\begin{table}[t]
\caption{\label{Table1}%
The vibrational parameters for oscillator fits to the infrared-active Fe-As mode observed in different doped compounds at 300~K and 150~K, where $\omega_{0}$, $\Omega$, and $\gamma$ are the oscillator frequency, strength, and line width, respectively. All units are in cm$^{-1}$.}
\begin{ruledtabular}
\begin{tabular}{ccccccc}
    &        & 300~K &    &      &  150~K &     \\
\cline{2-4}\cline{5-7}
Doping & $\omega_{0}$  & $\Omega$ & $\gamma$  &$\omega_{0}$  & $\Omega$ & $\gamma$ \\
\hline
BFA  & 253.10 & 211.85 & 3.77 & 258.00 & 217.13 & 3.43 \\
K40  & 251.30 & 321.50 & 5.51 & 255.88 & 322.74 & 4.51 \\
Co08 & 254.94 & 208.26 & 6.31 & 259.75 & 208.86 & 5.93 \\
P15  & 257.34 & 175.10 & 5.92 & 261.78 & 181.78 & 5.88 \\
\end{tabular}
\end{ruledtabular}
\end{table}

This brings us to the question of what causes the anomalous red shift of the phonon in K40. One possibility is that the phonon is coupled to other channels, such as charge or spin. In the quasi-two-dimensional quantum spin system Sr$_{1-x}$Ba$_{x}$Cu$_{2}$(BO$_{3}$)$_{2}$, strong spin-phonon coupling leads to an additional 3\% softening of the related phonon.\cite{Choi2003,Homes2009} The pronounced gate-induced phonon softening in bilayer graphene is attributed to the coupling of the phonon to electronic transitions.\cite{Kuzmenko2009} In addition, electron-phonon and spin-phonon coupling are widely observed in FeSCs.\cite{Mittal2009,Rahlenbeck2009,Choi2010,Chauviere2011,Niedziela2011} Theoretical calculations also point to spin-phonon coupling as the origin of the phonon softening in FeSCs.\cite{Huang2010,Li2011} This makes the electron- or spin-phonon coupling scenario more favorable to accounting for the anomalous phonon red shift in the K-doped compounds.

Further evidence for electron- or spin-phonon coupling can be revealed from the line shape of the phonon, since coupling of the lattice vibrations to charge or spin excitations often results in an asymmetric line shape.\cite{Fano1961,Kuzmenko2009} One may already notice that, as shown in Fig.~\ref{Fig2}(d), while a Lorentz oscillator yields a reasonably good description to the phonon line shape in BFA, Co08 and P15, the Lorentz fitting result for K40 is relatively poor: the phonon in K40 exhibits a slightly asymmetric line shape. We can emphasize the asymmetry of phonon line shapes by subtracting the background. The inset of Fig.~\ref{Fig4}(b) shows the phonon for all compounds, with the background removed from the total optical conductivity. The asymmetric line shape in K40 (blue solid circles) can be easily distinguished from other three materials, suggesting stronger electron- or spin-phonon coupling in the K-doped compound.

The asymmetry of the phonon can be quantified by fitting it to a Fano line shape:\cite{Fano1961}
\begin{equation}
\label{Fano}
\sigma_{1}(\omega)=\frac{2\pi}{Z_0} \frac{\Omega^2}{\gamma}
\frac{q^2 +\frac{4q (\omega - \omega_0)}{\gamma} -1}{q^2 (1 + \frac{4(\omega - \omega_0)^2}{\gamma^2})},
\end{equation}
where $\omega_{0}$, $\gamma$ and $\Omega$ represent the resonance frequency, line width and strength of the phonon, respectively. The asymmetry of the Fano line shape is described by a dimensionless parameter $1/q^2$. As $1/q^2$ increases, the asymmetry of the line shape intensifies, indicating a growth of the coupling strength. In the case $1/q^2 = 0$, the symmetric Lorentz line shape is fully recovered.
%%%%%%%%%%%%%%%%%%%%%%%%%%%%%%%%%
%Figure 4
The inset of Fig.~\ref{Fig4}(b) depicts the fitting results at 150~K for each material. The parameter $1/q^2$, determined from the Fano fit, is shown as green solid circles in Fig.~\ref{Fig4}(b). While the Fano parameter ($1/q^2 \approx 0.0016$) is vanishingly small in BFA, Co08 and P15, its value ($1/q^2 \approx$ 0.045) is almost 30 times larger in K40. This implies that the asymmetry of the phonon line shape is much stronger in K40, and thus the coupling between the phonon and electrons or spins is significantly enhanced in the K-doped compounds.

The coupling between lattice vibrations and electrons or spins could also cause an increase of the phonon intensity.\cite{Choi2003,Kuzmenko2009} The intensity of the phonon $\Omega^2$, determined from the Fano fit at 150~K, is traced out for all the compounds as blue solid triangles (error bars within the symbols) in Fig.~\ref{Fig4}(b). Comparing to BFA, $\Omega^2$ doubles in K40, but slightly decreases in Co08 and P15. The change of the phonon intensity (the area under the phonon line shape) can also be identified directly from the phonon line shape as shown in Fig.~\ref{Fig2}(c), Fig.~\ref{Fig2}(d) as well as the inset of Fig.~\ref{Fig4}(b). The slight decrease of $\Omega^2$ in Co08 and P15 may arise out of disorder effect induced by in-plane doping. The striking increase of the phonon intensity in K40 is supportive of coupling between the phonon and electrons or spins in the K-doped compounds.

One should note that our observations do not conflict with the evidence for the electron- or spin-phonon coupling in BaFe$_{2}$As$_{2}$,\cite{Niedziela2011} Co-doped BaFe$_{2}$As$_{2}$\cite{Chauviere2011} or other iron-pnictides,\cite{Mittal2009,Rahlenbeck2009,Choi2010} because our studies are based on the comparison between the parent compound BFA and doped materials. We are not saying that there is no coupling in BFA and Co- or P-doped materials. What we assert is that the coupling is stronger in the K-doped material when compared to the other compounds. The relatively stronger electron- or spin-phonon coupling might be an important factor for the relatively higher $T_c$ in the K-doped BaFe$_{2}$As$_{2}$ system.

%%%%%%%%%%%%%%%%%%%%%%%%%%%%%%%%%%%%%%%%%%%%%%%%%%%%%%%%%%%%%%%%%%%%%%%%%%%%%%%
%
% Conclusions
%

In summary, the effect of three different substitution types on the 253~\icm\ $ab$-plane infrared-active Fe-As phonon in \BFA\ has been studied by infrared optical spectroscopy. This mode shifts to higher frequencies in Co- and P-doped \BFA, in agreement with the expected frequency shift associated with lattice parameter change. Intriguingly, the same mode exhibits an unusual red shift in the K-doped compound. We attribute the anomalous red shift of the Fe-As mode in K-doped \BFA\ to electron- or spin-phonon coupling. The asymmetric line shape and intensity growth of this mode in K-doped \BFA\ also support our conclusion.

%%%%%%%%%%%%%%%%%%%%%%%%%%%%%%%%%%%%%%%%%%%%%%%%%%%%%%%%%%%%%%%%%%%%%%%%%%%%%%%
%
% Acknowledgment
%

We thank N. L. Wang for helpful discussion. Work at IOP CAS was supported by MOST (973 Projects No. 2012CB821403, 2011CBA00107, 2012CB921302 and 2015CB921303), and NSFC (Grants No. 11374345 and 91121004). Work at BNL was supported by the U.S. Department of Energy, Office of Basic Energy Sciences, Division of Materials Sciences and Engineering under Contract No. DE-SC0012704.

%%%%%%%%%%%%%%%%%%%%%%%%%%%%%%%%%%%%%%%%%%%%%%%%%%%%%%%%%%%%%%%%%%%%%%%%%%%%%%%
%
% The bibliography (BibTeX)
%

%\bibliography{biblio}

%

\end{document}